# Pulsed discharge plasma for graphite exfoliation in liquid nitrogen

DA SILVA TOUSCH C., LIEBGOTT Q., LETOFFE A., IBRAHIM D., KABBARA H., NOEL C., HENRION G., HEROLD C., ROYAUD I., PONCOT M., FONTANA S., CUYNET S.

IJL, Université de Lorraine, CNRS, 2 allée André Guinier, F-54000 Nancy, France

**Abstract:** Highly Oriented Pyrolytic Graphite was exfoliated via pulsed discharge plasma in liquid nitrogen. The potential mechanisms involved were investigated by observing the treated surface of the graphitic material and the obtained particles. Non-exfoliating defects from the plasma treatment were observed and experimental parameter were modified to counteract those. One experiment was performed without exposing the HOPG directly to the discharges so as to better understand the plasma's role. The exfoliated particles were observed via TEM and SEM to evaluate the defects, the size, the purity and the crystallinity but no quantitative characterization of their thickness was possible so the actual number of layer of each particle is unknown. Nonetheless, few layers graphene (FLG) was successfully exfoliated through this process. The proposed mechanisms were extrapolated from the observation of the damaged HOPG surface and the obtained particles but the correlation found does not prove causation.

## I. INTRODUCTION

As the first two-dimensional material discovered, Graphene with its single layer hexagonal structure began to be thoroughly researched since 2004. Back then, A. Geim and K. Novoselov of the University of Manchester purposefully mechanically exfoliated graphite pieces with the now famous duct tape method [1-3]. The growing interest for this honeycomb patterned sheet of carbon stems from its outstanding properties such as a thermal conductivity [4] and a mechanical strength [5] amongst the highest of all known materials. Furthermore, single layer graphene's theoretical specific surface area stands over 2500m$^2$/g [6] with a transparency up to 97% [7-8]. As far as electronic properties are concerned, graphene is a zero-overlap semimetal, allowing both electrons and holes to act as charge carriers [9], thus displaying a very high electrical conductivity. Given those properties, its potential applications reside within capacitors, batteries, sensors, transparent and/or flexible electronics, toxic-removal materials, etc..[7,10-12]

Various approaches for preparing graphene have been explored since; whether it be with a top-down or a bottom-up protocol. Mechanical exfoliation (including sonication), (electro)chemical exfoliation and oxidation-reduction methods consist in top-down approaches where a macroscopic carbonic material is degraded with the purpose of isolating graphene layers. Physical and chemical vapor deposition [13] are both bottom-up practices involving a carbonic precursor (such as methane for CVD) consumed to enable the growth of an atomic thin film on a substrate [19,20]. Each of those methods produce graphene of different quality, size and amount yet there is no perfect method offering high amounts of pristine large single layer graphene sheets to this day.

On the one hand CVD growth produces high quality graphene but the substrate used can contaminate it and the production rate and cost are not suited for industrial applications [14-17]. On the other hand, electrochemical exfoliation satisfies both production rate and cost requirements although the product obtained is of small lateral dimensions, contains a significant amount of defects/contaminations [18,25] and is more often than not few layer graphene instead of single layer [24], which properties are of a lesser magnitude.

Also, in the case of electrochemical exfoliation and oxidation-reduction methods, the obtained product cannot be qualified as graphene but instead graphene oxide (GO) [23,26]. This material is quite similar to graphene except for the presence functionalizing compounds such as hydroxyl, carboxyl and other oxidative functions. As such, GO dispersions become more stable in polar liquids and aggregation is reduced. Moreover, when such material is introduced in a polymeric matrix so as to obtain a composite material, the presence of such functionalizing compounds on the graphene sheets allows for better compatibility between the GO and the matrix resulting in stronger matrix-reinforcement interface and better dispersion and distribution of the GO in the matrix. Nonetheless, the electronic structure of graphene is disrupted by such functionalization making GO an insulator instead of a semi-metal. In fact, the overall properties of graphene oxide are substantially different from graphene and as such the distinction is crucial.

More recently, graphene and other carbon nanostructures have also been synthesized through low temperature plasma assisted CVD (PACVD) and arc discharge. In PACVD, the main challenge is to deliver the correct amount of carbonic precursor to sustain a rapid uniform nucleation all the while avoiding or limiting damages from the ion bombardment [19,20]. As for the arc discharge method, it entails using a pulsed discharge plasma as a mean of exfoliating graphene flakes from a bulk carbon material submerged in a liquid [21-23]. The carbon material is either directly used as an electrode or put in the liquid. Different carbon nanostructures were obtained through such method [21].

The guiding idea rests on the anisotropic nature of the different bonding interactions in graphite. Graphite can be viewed as a large stack of graphene layers. Within one layer, carbon atoms adopt a $sp^2$ orbital hybridization and as such, neighbor atoms are held together by strong covalent bonds. However, two adjacent planes are kept together only by van der Waals forces, which in comparison are of a much lesser strength. This difference between intra-plane and inter-plane cohesion is crucial to being able to separate layers without damaging their integrity. With a controlled energy, it is possible to overcome the van der Waals forces without destroying the covalent C-C bonds within the layers.

In this paper we explore the exfoliation of highly orientated pyrolytic graphite (HOPG) through arc discharge plasma in liquid nitrogen. This solution plasma process (SPP) aims at benefitting from a moderately low-energy plasma to exfoliate graphene flakes without damaging the carbon honeycomb structure. Liquid nitrogen is preferred over water as to not oxidize the produced graphene flakes. Had water been used, the reactive species in plasma from water molecules could react with the graphene flakes and bind oxidizing functions onto it. Such cold plasma process has been reported as an effective method to synthetize nanoparticles from metallic electrodes. [30]

## II. MATERIAL & METHODS

Highly Oriented Pyrolytic Graphite (HOPG) plates are used as electrodes and carbon precursor. This type of synthetic graphite is polycrystalline whilst maintaining good alignment, especially in the [0001] direction (Miller-Bravais hexagonal lattice coordinate system) of the hexagonal close packed structure of graphite. Different respective orientations of the electrodes were tested while maintaining a constant 0.5mm inter-electrode gap during the experiment. The discharge between the two electrodes was generated in a cryogenic Dewar vase in order to slow down the vaporization of the liquid nitrogen. A schematic representation of the experimental setup is shown in Figure 1. The High Voltage Power Supply operated at different voltage between 5 kV and 10 kV. The waveform function generator coupled with a high voltage switch allowed working with frequencies ranging from 10 Hz to $10^4$ Hz. Different pulse durations were tested: from 200 ns to $10^{-2}$ s. Voltage was measured by a high voltage probe and the different signals were observed on an oscilloscope [LeCroy Wavesurfer 104 MXs-B]. To get a more representative perception of the obtained particles, no resuspension nor sonication were performed. Instead, a polished silicon wafer, upon which TEM grids were deposited, was placed at the bottom of the Dewar container before the experiments. Varying experiment durations were explored as well ranging from 15 seconds to 20 minutes.

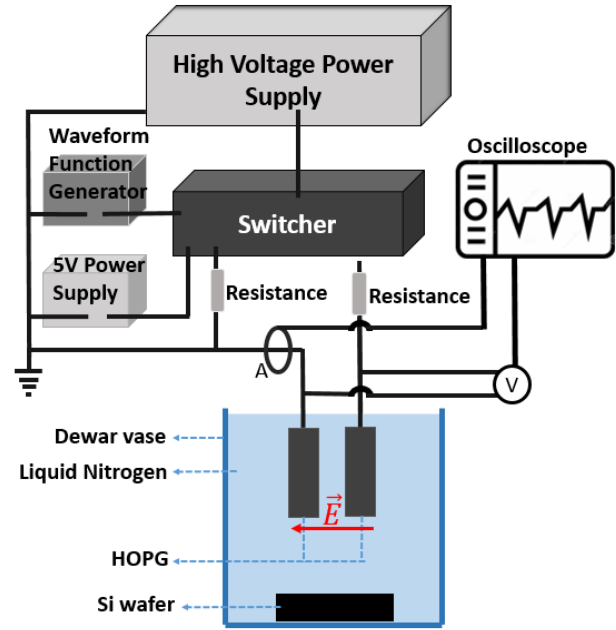

Figure 1: Schematic representation of the experimental setup.

Although electrical characterization would have been an interesting matter, given the pulse width of about 300ns, the data recorded by the oscilloscope were not exploitable. Indeed, at such small time scales, with great variations of the voltage and current during the discharge, the readings on the oscilloscope are inaccurate. The only use of the current and voltage measurements was to confirm the frequency, pulse width and applied voltage before each experiment and also to indicate the point of contact between the two electrodes by applying a low voltage and checking whether or not current was null.

As for the electrode respective orientation and position, 5 different setups were tested. As shown in Figure 2, the configuration chosen were the result of the combination of technical limitation and careful thinking as to how exfoliation could be optimized.

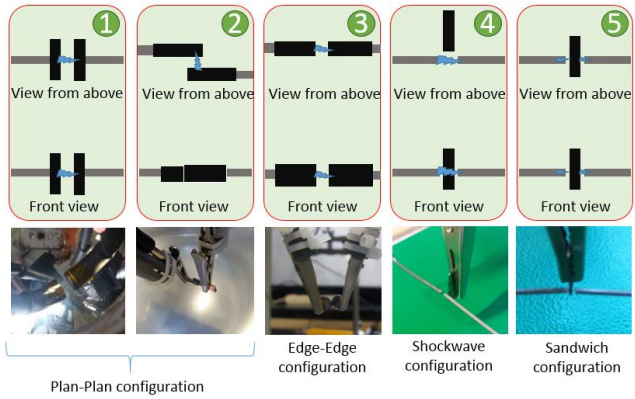

**Figure 2: Schematic representation and pictures of the relative electrode orientation.**

The fixation mechanism was at first the limiting factor for the duration of the experiment. In set-up 1, 8mm diameter HOPG discs were glued on brass 8mm rods with silver lacquer so as to bind them while maintaining electrical conductivity. The brass rod is insulated with a thermoretractable sheath and Kapton tape surrounds the junction between the HOPG and the rod to prevent arcing from the metallic electrodes. This fixation with silver lacquer held up for a few minutes but one of the HOPG discs fell off and it was decided to switch to a mechanical clipping system as seen on Figure 2 for all set-ups except n°1.

Regardless of the set-up configuration, the exfoliated particles were recovered for characterization by placing a polished Silicon wafer and TEM grids in the Deware during the experiment and letting the liquid nitrogen completely evaporate before collecting them. No resuspension, centrifugation nor sonication was performed and the Si wafer and TEM grids were observed directly.

The HOPG electrodes and the obtained particles were examined through scanning electron microscopy (SEM) [FEG Environmental FEI QUANTA 250, 15-20kV acceleration voltage] to assess the morphology of the coarser particles and the compared surface aspect of the electrodes before and after treatment. Samples for SEM characterization consisted in the polished silicon wafers recovered after the experiments once the liquid nitrogen had completely vaporized. As for the electrode they were simply recovered and examined with no intermediate treatment either. During those observation, BSED was used to detect contaminations from the experimental set-up, on the carbon flakes. Also the aim in examining the electrodes was to point out the different possible exfoliation mechanisms and the defects resulting from the experiment. By no means was it possible to quantitatively evaluate the thickness of the obtained flakes, however, backscattered electron imagery allowed a qualitative evaluation of said thickness by comparing the relative transparency of different particles.

A more precise evaluation of the smaller particles collected on the amorphous carbon-Cu grid was performed via transmission electron microscopy (TEM) along with Selected Area Electron Diffraction (SAED) [Philips CM200, $LaB_6$ electron cannon, 200kV acceleration voltage]. During TEM observation, the morphology and defects of the flakes, as well as their crystalline quality, were investigated. Indeed, the diffraction patterns allowed the differentiation of monocrystalline, polycrystalline and amorphous particles. This study was coupled with Electron Diffraction Spectroscopy (EDS) to confirm the nature of the observed objects. Two different kind of TEM grids were used: at first plain amorphous carbon-Cu grid and then holey amorphous carbon-Cu grid. While no major issue was encountered with the former one, it was opted to switch to holey grids given the thinness of the deposited particles. The initial purpose was to isolate the edge of a well oriented graphene flake count the number of layers to deduce the overall thickness given the interplanar distance, but it was to no avail because of technical limitation from the microscope.

Optical spectroscopy would have been a useful plasma diagnosis characterization method, however, the complexity of the set-up and the fact that the plasma is generated in liquid nitrogen prevented the use of such method.

### III. RESULTS

III.a. DEGRADATION OF THE ELECTRODES

Periodic readjustment of the inter-electrode distance to maintain the pulsed discharge regimen was necessary. It indicates that the degradation of the electrodes was fast enough to increase gap size to a value where the voltage difference between each electrode was too low to induce a discharge according to Figure 3 [27]. Indeed, in a pulsed regimen where the dielectric is self-healing and the inactive time is greater than the time necessary for the liquid medium to revert back to its initial state (prior to the first discharge), each electrical breakdown is performed in the same conditions of pressure, temperature and inter-electrode medium. For discharge frequencies between 10 Hz to $10^4$ Hz with pulsed duration of $10^{-6}$ s, each electrical breakdown occurs with the same initial conditions: the temperature, pressure and inter-electrode medium (liquid nitrogen) are the same. As such only the inter-electrode spacing will dictate the value of the breakdown voltage. As shown on Figure 3, as spacing increases, so does the breakdown voltage. When the gap becomes so wide that the breakdown voltage is greater than the applied voltage, no discharge occurs and the inter-electrode distance must be readjusted in order to reignite another discharge.

This erosion of the electrode is confirmed by visual cues during the experiment: some visible particles are projected in the liquid following the discharge.

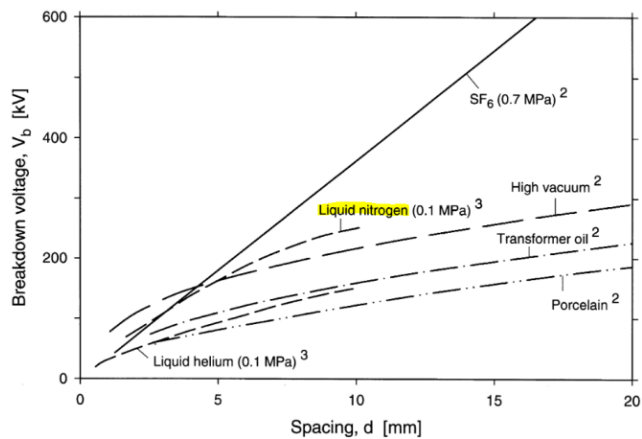

**Figure 3: Breakdown voltage as a function of inter-electrode spacing for liquid N₂ and other solid and liquid medium. [38]**

### III.a.1. PLAN-PLAN CONFIGURATION

Those results are similar for set-up 1 and 2 where the electrodes' flat surface are faced to one another.

The observed HOPG plates were subjected to a 20 s experiment with applied voltage 6 kV, frequency 10 Hz and pulse duration 0.05 s. The experiment stopped when one of the electrodes broke in two pieces. Analysis of the electrode surface through scanning electron microscopy revealed different types of degradation linked to potential exfoliation mechanisms. A plain comparison of the surface exposed to the plasma of the electrode before and after treatment confirms that there is indeed an exfoliation of particles. As shown on Figure 4, the exfoliation process manifests at different scales, producing particles with radically different lateral dimensions and thickness. Overall, the thickness is proportional to the lateral dimensions of the exfoliated particle: smaller particles were the thinnest.

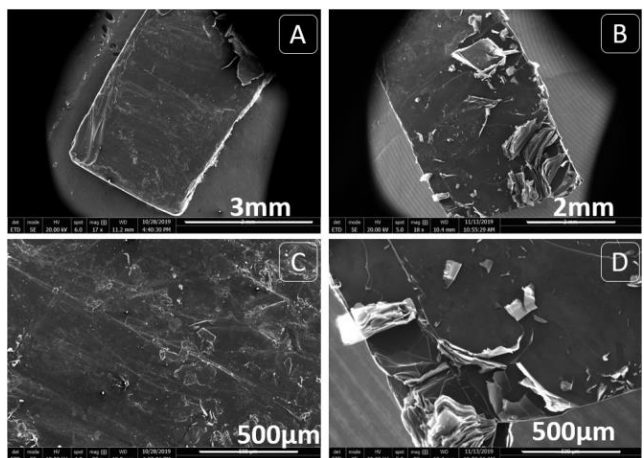

**Figure 4: Surface aspect of the HOPG plate from the 2nd set-up before (A and C) and after (B and C) treatment for 1 minute.**

As shown on Figure 5, the layers are torn from the electrode revealing the underlying planes then exposed themselves to the plasma discharges and torn in a similar way. Also, some weak points on the electrode where stress concentration is likely to occur, are significantly more impacted: in particular, the sharp corners of the rectangular HOPG plate from set-up 2 as seen on Figure 4 and 5.

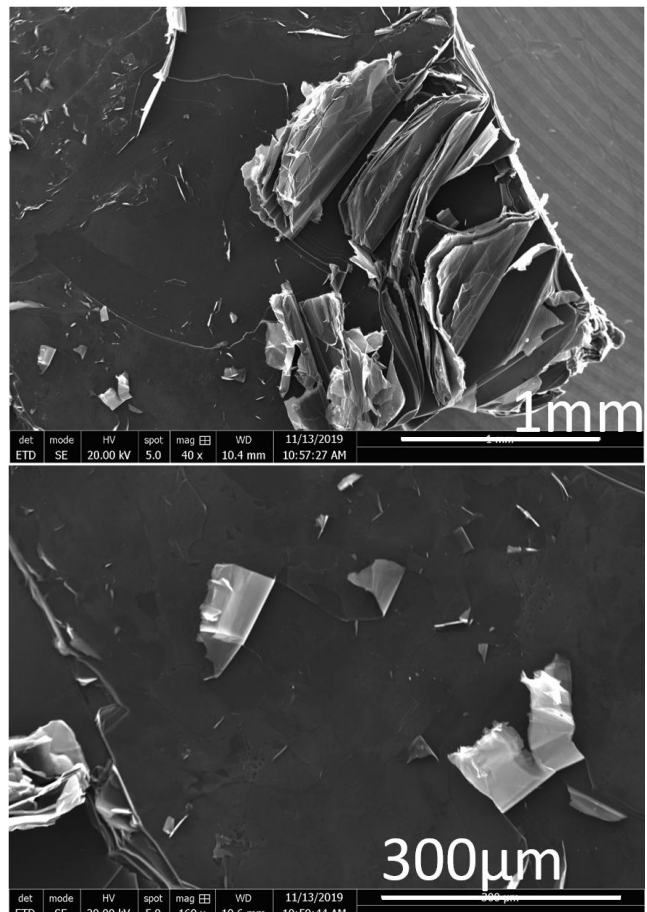

**Figure 5: Exfoliation in plan-plan configuration from set-up 2.**

In parallel, counterproductive degradations of the HOPG were noticed. Those are only damaging the electrodes which leads to damaged exfoliated particles. Such defects can be seen on Figure 6 and at that point it is important to distinguish the positively polarized electrode from the negatively polarized one. Indeed, on the positively polarized electrode the defects consist in circular smooth-edged holes with varying diameter ranging from 0.5 µm to 4 µm (Figure 6 A) and a cracked top layer (Figure 6 B).

It is supposed that the holes are the result of arc roots that very locally sublimate the carbon from the electrode leaving a punctual defect on the surface (anodic spot). It was not possible to assess the depth of those holes but it is certainly affecting a lot of carbon layers. As far as the cracked top layer is concerned, this defect was only observed on the outmost layers and more widespread than the regrouped punctual holes. It does not appear to be mechanical failure of the layer but rather a superficial coating deposited on the surface that would have cracked upon quenching. Indeed, the local sublimation of carbon during the discharge could have induced a subsequent deposition and solidification on the surface after the discharge had stopped. Several explanations can be considered: this deposited layer could be rapidly cooled by the liquid nitrogen resulting in high thermomechanical stresses on this thin layer and giving this cracked aspect; or it could be right before observation in the SEM's vacuum chamber that small imperfection containing trapped gas burst out when their inner pressure exceeds the outer pressure [39].

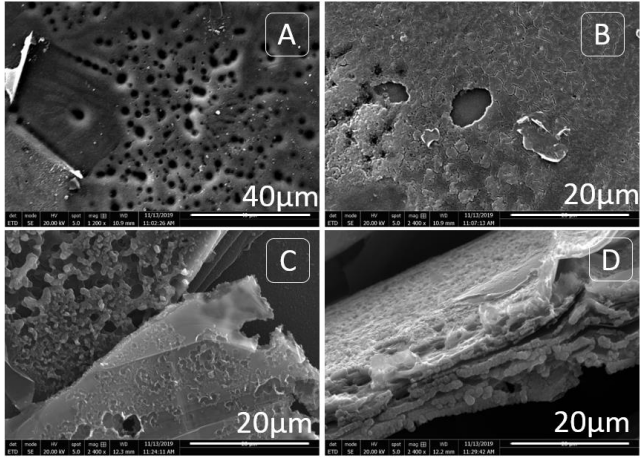

**Figure 6: Defects after plasma treatment on positively polarized HOPG electrode (A&B) and ground HOPG electrode (C&D).**

On the negatively charged HOPG electrode the observed defects can be described as shreds in thin layers ( Figure 6 C, bottom right ), spheroidal alveoli-like structures ( Figure 6C, top left, and 6D ) and a very rough surface aspect on the outer layer ( Figure 6D ). The shreds are dispersed throughout the entire electrode on the outermost layers and especially on partially exfoliated particles. They are not similar to the smooth holes described earlier in the way that they do not appear as deep and they have a rough contour indicating that it has been torn apart or that it has exploded rather than being sublimated. The spheroidal cauliflower structures have been confirmed to be composed of carbon through EDS and the aspects leads towards being bulk recondensed amorphous carbon. During the discharge, rapid temperature elevation ensues and temperature above 5000K are reached thus potentially sublimating carbon. After the discharge stops, liquid nitrogen floods the inter-electrode gap and quenches any residual vaporized carbon onto the electrode surface. This repeated phenomenon could explain the formation of such amorphous, disorganized structures on the surface and the exposed plan edges of the electrode. Concerning the rough surface aspect, it appears to be a combination of shreds and amorphous carbon deposition on the electrode.

### III.a.2. EDGE-EDGE CONFIGURATION

Those HOPG plates were subjected to a 5 minutes experiment with applied voltage 6 kV, frequency 500 Hz and pulse duration 1 µs

In this configuration, damages to the electrode were visibly greater than in the previous set-up, as shown on Figure 7 (A and B). Once again, the layer separation is manifested both macroscopically and microscopically. Figure 7 (C and D) shows how the HOPG was split open by the plasma treatment and two main areas can be distinguished. On both micrographs, the right side shows very coarse and thick layers being separated and peeled of the HOPG disc and the left side shows a similar exfoliation process but at a much smaller scale with thinner and smaller flakes being peeled off. It should be noted though, that while the previous set-up held-on for a minute before an electrode broke, this experiment lasted for more than 10 minutes so direct comparison are not appropriated.

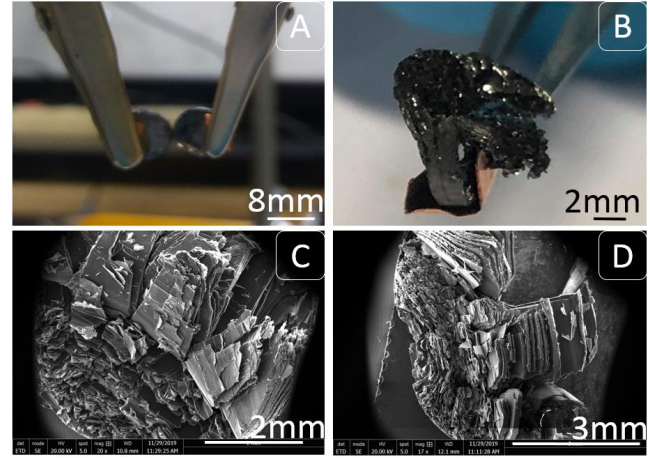

**Figure 7: HOPG electrodes before (A) and after plasma treatment (B). SEM images of the electrodes after plasma treatment (C): ground electrode; (D): positively polarized electrode.**

As far as the small scale exfoliation is concerned, non-exfoliating defects were also spotted. Unlike the previous set-up, no smooth holes nor amorphous looking carbon were spotted on either electrode as shown on Figure 8.

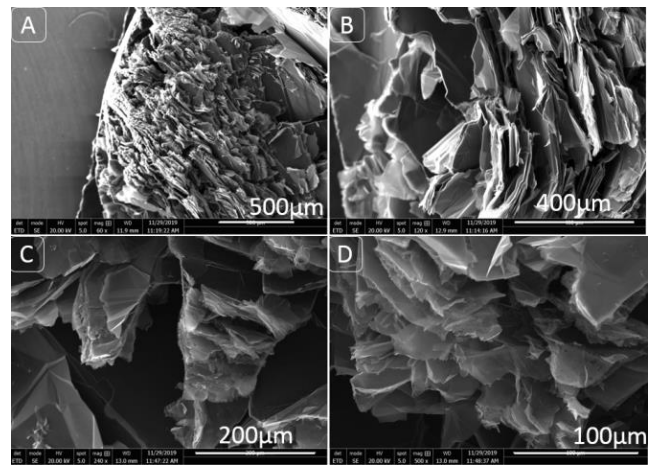

**Figure 8: SEM images of the small scale exfoliation of the HOPG electrodes with increasing magnification. (A) and (B): positively polarized electrode. (C) and (D): ground electrode.**

Nevertheless, on the negatively polarized electrode, punctual defects similar to the ones spotted on the same electrode from the previous set-up were spotted as shown on Figure 9. Those are hypothesized to be formed when streamers escape this negatively polarized electrode in the first instants of the pulse.

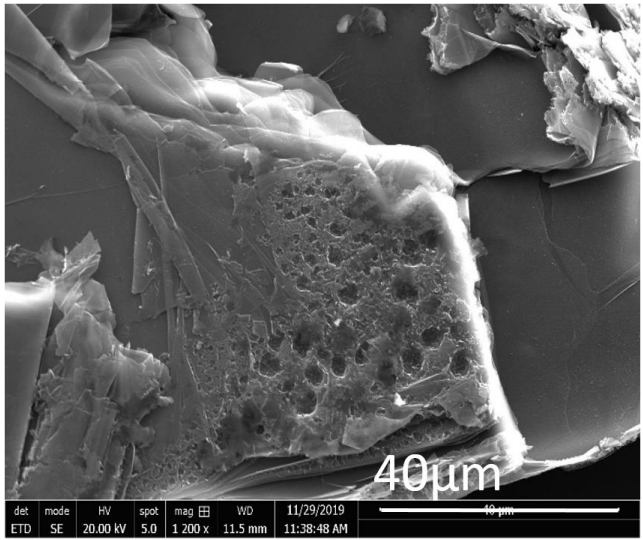

**Figure 9: Non-exfoliating punctual defects on the ground HOPG electrode.**

### III.a.3. SHOCKWAVE CONFIGURATION

In this set-up, the electrode is not directly exposed to the plasma as it does not act as an electrode. Instead, two tungsten electrodes are used for the arc discharges and the HOPG plate's edge is held 0.5mm away from the inter-electrode gap, perpendicularly to the tungsten electrodes. In that way, the shockwaves induced by the arc formation impact the HOPG but the temperature elevation and the arc itself do not affect the graphite. For this set-up, the parameters were: applied voltage 6 kV, frequency: 500Hz, pulse: 300ns different process durations were tested to assess evolution of the process through time. This experiment ended up being less destructive for the electrode for small duration (5minute), however, for longer durations (10 and 20 minutes), macroscopic degradation was noticeable as shown on Figure 10.

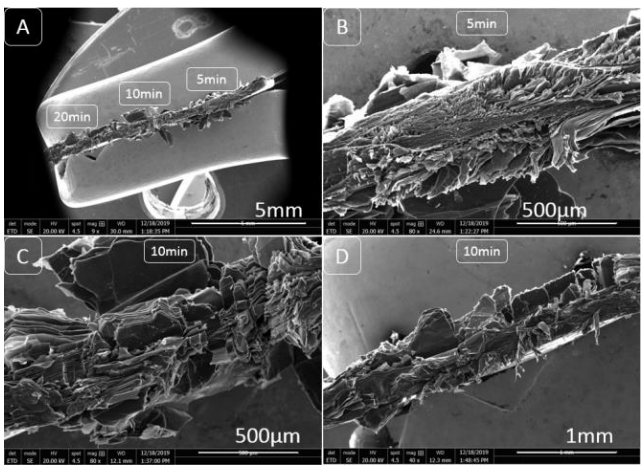

**Figure 10: SEM images of the degraded state of the HOPG plate after shockwave treatment of increasing duration.**

As far as process time is concerned, there is a significant difference between 5 minutes and 10 minutes, however 10 minutes and 20 minutes treatment offer very similar results. It should be noted that the HOPG plate was not moved during the experiment and remained at its initial distance from the electrodes.

On the one hand, non-exfoliating defects were not detected and no big particles were projected in the liquid during the experiment. It seems that this approach is indeed much less destructive but as a consequence, the exfoliation rate is severely reduced.

On the other hand, tungsten contaminations were spotted via BSED imaging and confirmed through EDS. Those contaminations come from the degradation of the electrodes during the process.

### III.a.3. SANDWICH CONFIGURATION

In this set-up, the HOPG plate is sandwiched between the two tungsten electrodes while maintaining a 10µm gap between each part. The HOPG plate is not connected to the electric circuit and has a floating potential. As a consequence, when a voltage difference is applied between the two electrodes, the HOPG plate is subjected to the electric field induced by the voltage difference between the positive and negative electrodes. It provokes a polarization of the HOPG plate. When the induced voltage difference reaches the breakdown threshold of the system, two arc discharges could occur between each electrode and the two sides of the HOPG plate. This experiment was performed at applied voltage 5kV, frequency 500Hz and pulse duration 300ns. A piece of the HOPG broke after 4 minutes, marking the end of the experiment.

Figure 11 shows the damage caused to the HOPG plate after four minutes of process. There is a hole in the electrode where it was surrounded by the tungsten electrode and it looks like the HOPG plate exploded at that point. Visible particles were projected at a fast rate, in the liquid, during the experiment.

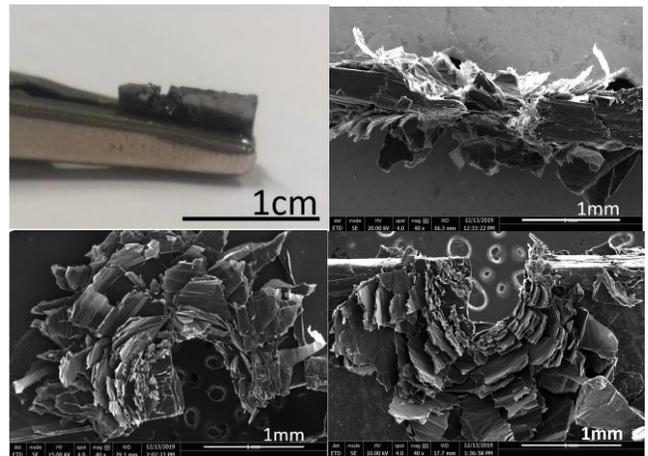

**Figure 11: Damaged HOPG plate from sandwich configuration experiment (A), SEM image from edge perspective (B), SEM image form left perspective (C), SEM image from right perspective (D).**

With this set-up, once again exfoliation is seen at different scales. Non-exfoliating defects were found and tungsten contaminations were also spotted and confirmed via spectroscopy analysis. Those defects and contaminations are shown respectively on Figure 12 and Figure 13.

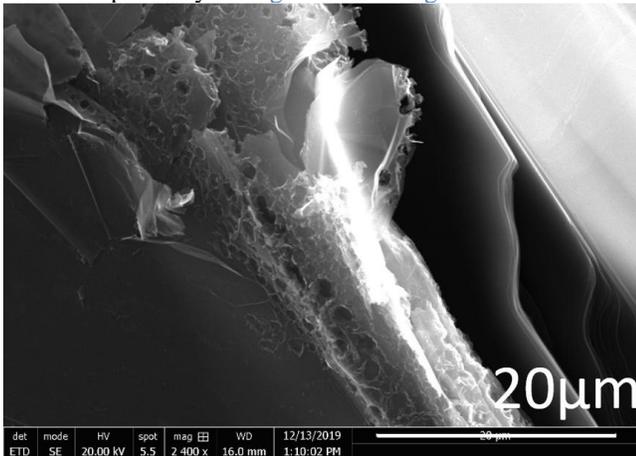

**Figure 12: Non exfoliating defect on HOPG plate used in sandwich configuration.**

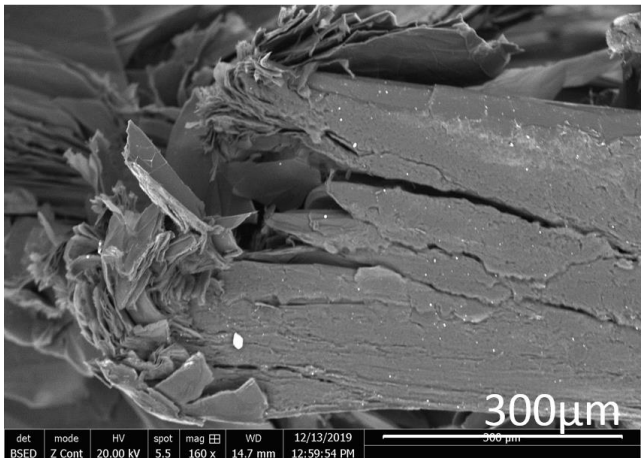

**Figure 13: Tungsten contaminations on HOPG plate used in sandwich configuration. BSED imaging, revealing white spots confirmed to be tungsten particles.**

### III.b. EXFOLIATED PARTICLES

The results from set-up 1, 2 and 3 have to be separated from those of set-up 4 and setup 5 since the HOPG is not used as an electrode in those two later ones.

#### III.b.1. HOPG AS ELECTRODES

Overall, particles obtained had lateral dimensions up to 500µm and down to a few µm. At first, SEM observations shown on Figure 14, indicated that the obtained flakes' thicknesses were not homogeneous and not always proportional to the lateral dimensions: on Figure 14 A, the flake has large lateral dimensions and appears almost transparent which indicates its relative thinness while on Figure 14 B, the particle is apparently much thicker while having small lateral dimensions.

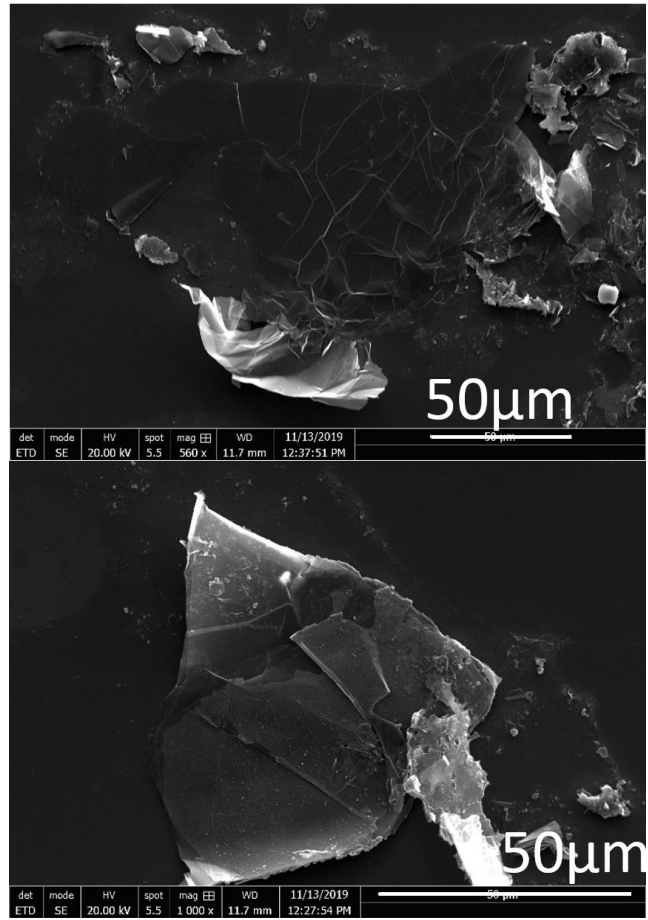

**Figure 14: SEM images of the particles from PLAN-PLAN configuration. (Top) Wider and thinner particle. (Bottom) Smaller and thicker particle.**

Also, as demonstrated on Figure 15 A, the defects spotted on the electrode were present on some particles and amorphous looking carbon particles were also found (Figure 15 B).

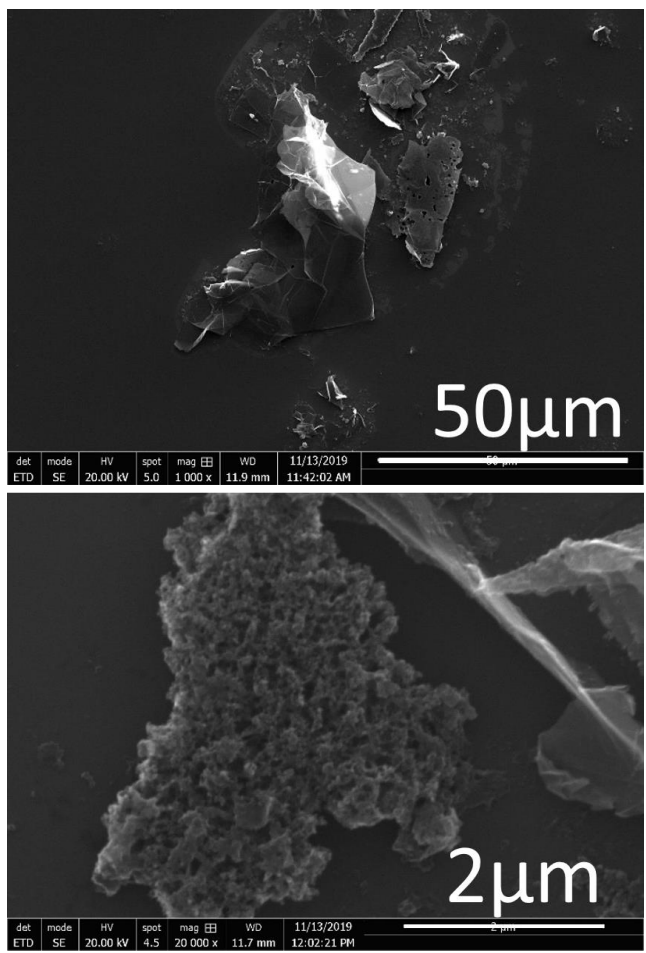

**Figure 15: SEM images of the damaged particles from PLAN-PLAN configuration. (Top)Holey particle. (Bottom) Amorphous carbon particle.**

TEM characterization confirmed the amorphous nature of those via analyzing the diffraction pattern presented in Figure 16. Although some of the particles were damage-free and contamination-free, most were contaminated by amorphous carbon, especially on the edge of the flakes (Figure 17 A). The dimensions of the holes in the holey particles were compared to the ones on the electrodes to find that they were identical. In general, there was a high diversity in the particles obtain both in terms of dimensions and in term of quality and defects. The presence of amorphous carbon is unwanted and is what actually lead to a change in the set-up.

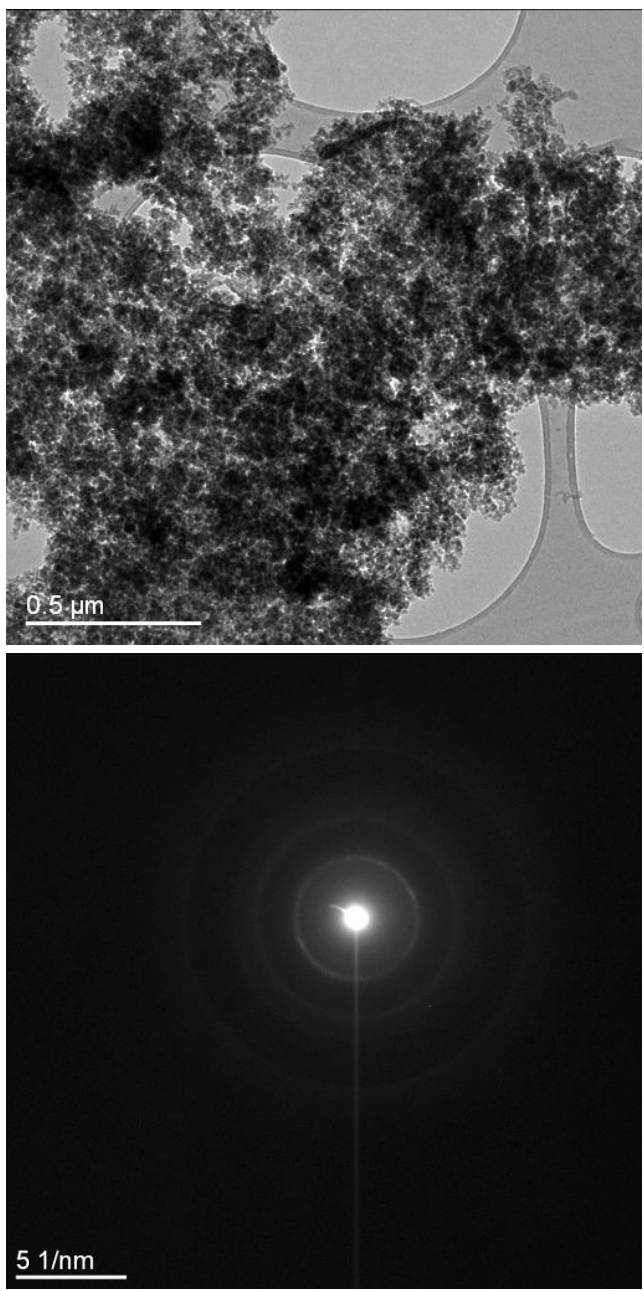

**Figure 16: TEM characterization of amorphous carbon. (Top) Amorphous particle. (Bottom) Associated diffraction pattern.**

### III.b.2. SHOCKWAVE CONFIGURATION

For this experiment, even with process time up to 20 minutes, the quantity of exfoliated particles was significantly inferior to the other configurations, although the HOPG plates were visibly degraded, SEM investigation revealed the presence of thicker and less qualitative flakes when compared to the other processes. One major improvement though is the absence of any structural defects nor amorphous carbon but instead tungsten contamination was omnipresent as confirmed via spectroscopy. Also, TEM imaging revealed that some graphene flakes were warped around tungsten nanoparticles,

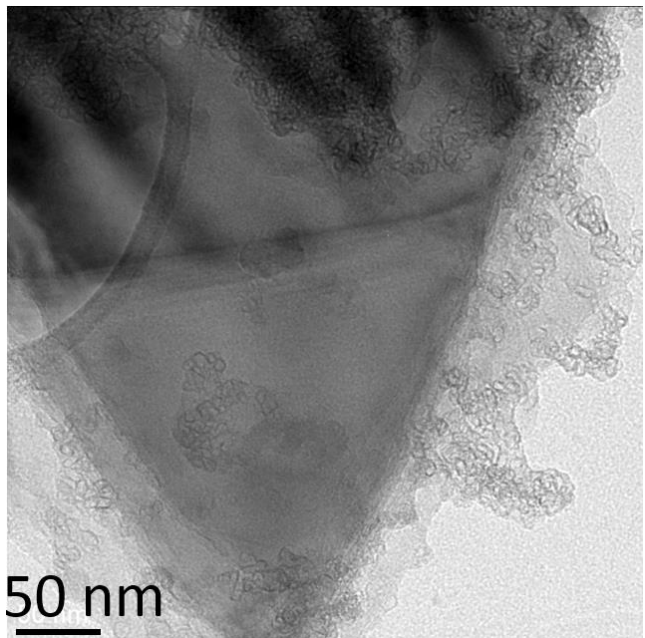

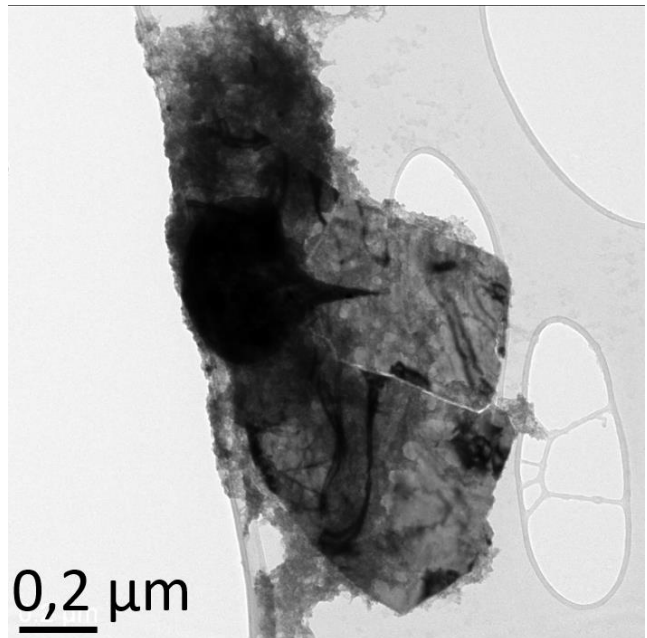

Figure 18: TEM image of a tungsten nanoparticle wrapped in a graphene flake from the shockwave configuration.

### III.b.3. SANDWICH CONFIGURATION

Exfoliated particles were more numerous in this set-up when compared to the previous one. Although some structural defects were spotted during SEM investigation, it only concerned a minority of the observed particles and the majority was seemingly defects-free. A recurring problem however, was the presence of tungsten contaminations, from the electrodes, on some particles. Neither SEM nor TEM observations showed amorphous carbon particles and only a very few graphene flakes observed via TEM displayed amorphous looking carbon on their edges. Almost completely transparent flakes were found (Figure 19) and overall, the crystalline nature of most flakes was monocrystalline or oftentimes it hinted at a turbostratic stacking of monocrystalline layers (Figure 20).

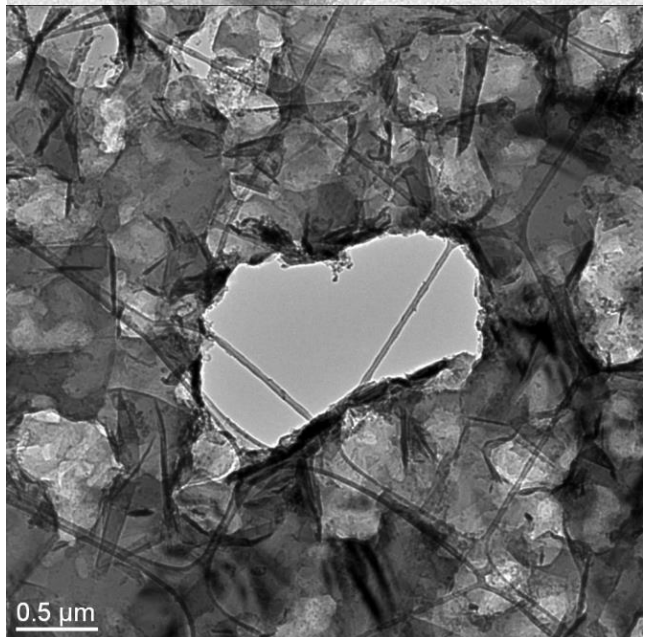

Figure 17: TEM micrographs of damaged particles. (A) Contaminated edge of a few layer graphene flake. (B) Holey particle.

as shown on Figure 18. Although, only a very few particles were in this configuration. As mentioned before, there was no significant difference between the surface aspect of the HOPG in the 10 minutes and 20 minutes experiment. Consequently, as far as the obtained particles are concerned, the same conclusion was reached. For the 5 minutes experiment however, the particles were especially sparse.

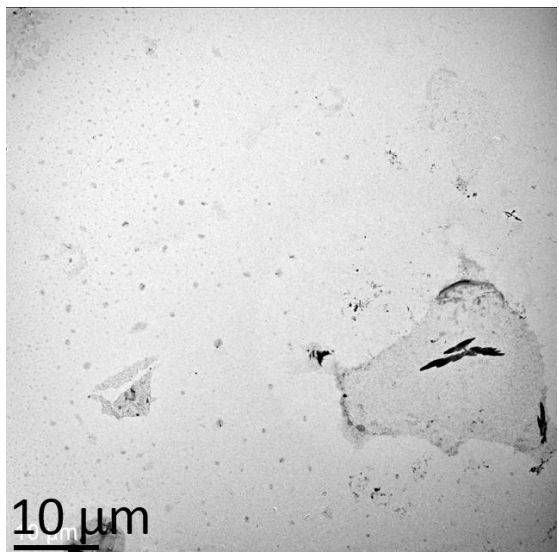

**Figure 19: TEM image of high transparency graphene flakes from the sandwich configuration experiment. Dark spots are tungsten contaminations.**

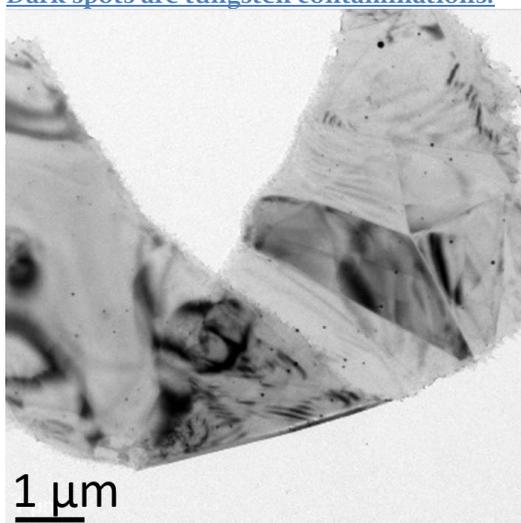

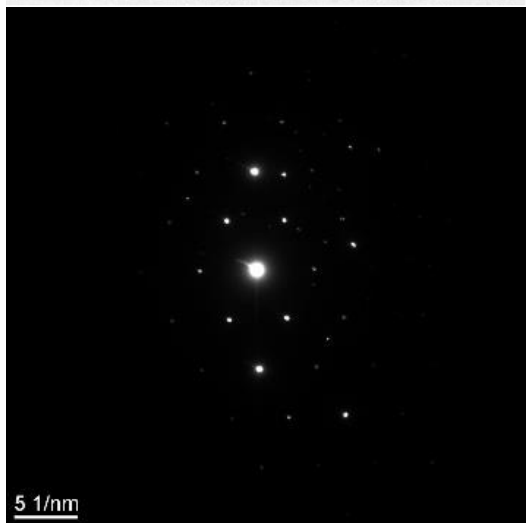

**Figure 20: TEM image and corresponding diffraction image of a graphene flake from the sandwich configuration experiment. Black dots are tungsten nanoparticles.**

## IV. DISCUSSION

As far as the exfoliation mechanisms are concerned, there are many parameters involved. Indeed, liquid pulsed discharge plasma is a two-step process where the pre-breakdown and post-breakdown period are distinguished. According to previous studies, [28-31] pre-breakdown the voltage oscillates before it drops marking the breakdown itself. During this period, a small ohmic current is generated then followed by the dissociation of the liquid, the, electrons are ripped from the cathode and only then does the breakdown occur. Although the details of this process depend a lot on the exact composition and properties of the liquid medium, those general steps apply [32-34]. (Figure 21)

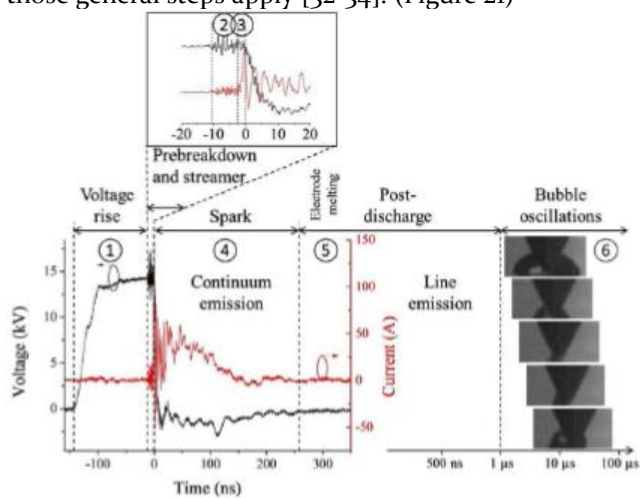

**Figure 21: Steps involved in the generation of a discharge in a dielectric liquid. (1) Voltage increase without charge injection. (2) Pre-breakdown: beginning of charge injection in the liquid. (3) Streamer propagation and sudden current increase. (4) Transition to arc discharge characterized by voltage drop and accompanied with intense UV and visible light emissions. (5) Post-breakdown and discrete light emissions. (6) Oscillations of the cavitation bubble formed by the discharge.** [33]

To this day there is no unanimously agreed upon predictive model describing the breakdown itself. Two major hypothesis stand: one being that streamers are introduced in the liquid itself and that the breakdown occurs in the liquid and the other being that the liquid transitions to gas before any discharge occurs [30].

The interaction mechanisms between the discharge and the electrodes are complex, intertwined and not precisely known. Indeed, during each discharge, energy is exchange between the plasma and the electrode, the temperature of the electrodes increases due to species and could be eroded, as in plasma sputtering process. The duration of the discharge is thus a key factor in controlling the erosion of the electrodes. In principle, the longer the pulse, the more energy is given to the electrodes and the greater the volume eroded; notwithstanding, the power supply may, in some cases, be a limiting factor as far as how much energy is brought to the electrodes. As such, depending on the properties of the generator, longer discharge duration is not always accompanied by a greater eroded volume.

When the electric field is applied, streamers could propagate in the liquid at speeds on the order of $10^5$ m/s which is several times higher than the speed of sound in any medium [35]. This can generate successive shockwaves impacting on both electrodes and inducing mechanical stresses. This streamer is most probably emitted from an irregularity on the electrode's surface that gets eroded and vaporized or explodes. Once the streamer reaches the cathodes, a conductive canal is formed and the voltage difference drops while the current intensity increases drastically. This is accompanied by light emissions in the IR, UV and visible spectrum depending on the exited elements [30]. The medium heats up and the surrounding liquid nitrogen vaporizes, forming a growing gas bubble.

After the discharge, the voltage and current drop to 0 V and 0 A, the emitted light fades and the electrodes start cooling down. In the interelectrode gap, a nitrogen gas bubble remains [36]. The surrounding liquid's pressure makes the bubble dilate and collapse, resulting in oscillations of the surface of the bubble while it is being evacuated from the inter-electrode gap. This phenomenon has been described by Rayleigh-Plesset (1949) and Gilmore (1952) models. Finally, the system reverts back to equilibrium: the electrodes are quenched by the liquid and the inter-electrode gap is in the same conditions as prior to the previous discharge. However, after each discharge, the electrode's surface is altered because of the erosion and the gap increases gradually so it is important to continually adjust the electrode's distance to pursue the experiment in the same conditions.

The observation of the electrode's defect further confirms some of the previously described model's mechanics. Indeed, the amorphous carbon and shredded layers on the ground electrode (Figure 6 C and D) seem to be the result of streamer emissions and drastic temperature increase, over the vaporing temperature of graphite. This carbon vapor could be immediately quenched post-discharge when liquid nitrogen floods the electrode's surface and carbon may be rapidly cooled on the surface of the electrode. Because of the great cooling rate, crystallization does not occur and the carbon solidifies in an amorphous state. As far as the positively polarized electrode is concerned, the holes from Figure 6 A appear to be anodic spots marking the point where the conductive canals are formed [37]. It is hypothesized that the reason why no amorphous carbon is spotted on the anode is possibly because the vaporized carbon is subsequently ionized and attracted to the cathode because of the polarization and the applied electric field as schematized in Figure 22 below.

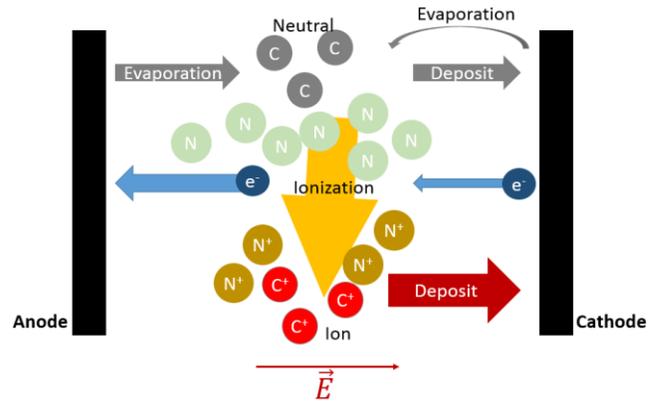

Figure 22: Schematic representation of the evaporation and deposition of carbon on the HOPG electrodes during the discharge.

Another phenomenon that can occur with such set-up is gas absorption at the cathode. As shown on Figure 23 below, at some point, the surface of the cathode shows erupted or collapsed bubbles patterns indicating that indeed some species were absorbed below a certain number of layer, they formed bubbles below the surface that stretched the material on top and then ended up erupting or collapsing on themselves as the gas evacuated.

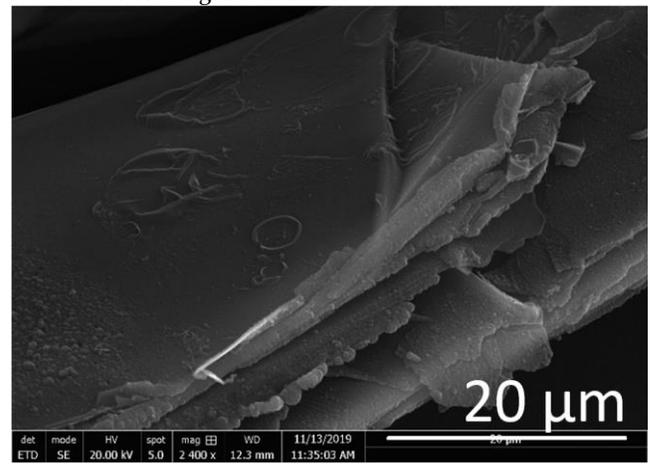

Figure 23: SEM image of erupted/collapsed bubble patterns on the surface of the HOPG used as cathode in set-up 2.

As far as the exfoliation itself is concerned, there are a lot of potential factors that come into play. The plasma erosion itself seems more destructive than actually the main exfoliating mechanism, however, the greatly reduced exfoliation rate of the shockwave experiment suggests that it is still a factor encouraging the exfoliation. The shockwaves generated from the discharge seem sufficient to exfoliate particles from the HOPG but the obtained particles are relatively thick (large stack of layers) and have no structural defects. As such it can be supposed that the plasma is necessary to separate them into few layer graphene(FLG).

The wear and tear caused by the discharges on the electrodes may be the initiator of the layer separation by generating stress concentration sites. Once the layers are locally degraded and separated, the stresses from the shockwave and from the cyclic increasing and decreasing of the temperature are enough to tear a few layers and thus exfoliate some particles. Also, the gas absorption at the cathode and the eruption of bubbles formed below a few layers are potential factors contributing to the exfoliation process.

Although some particles contain those same defects spotted on the HOPG electrodes, there are also damage-free particles. Because of that observation, one can assume that a first discharge is powerful enough to damage a few layers on the surface of the electrode and initiate exfoliation on this precise spot but on the subsequent discharge, most likely occurring elsewhere on the surface of the electrode, the area affected by the first discharge is still subjected to shockwaves and thermomechanical stress that pursue the exfoliation process on non-damaged layers (Figure 24).

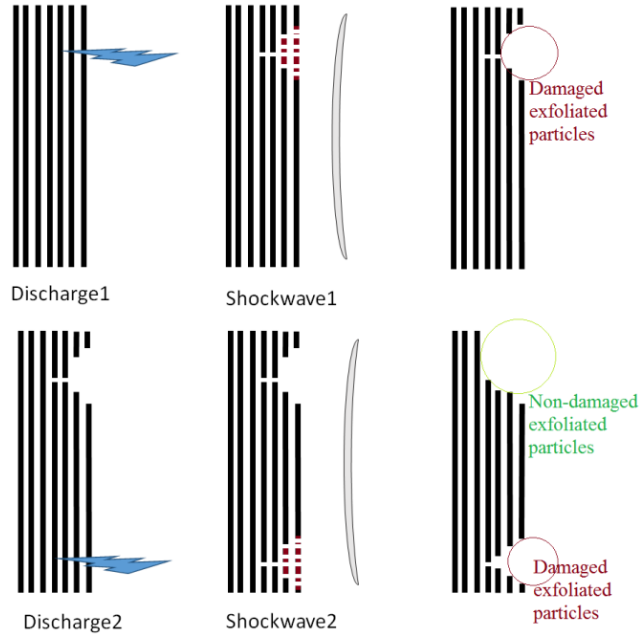

**Figure 24: Hypothetical schematic, sequential events describing the suggested exfoliation process.**

Also, in the plan-plan configuration, discharge duration was experimented with. Two experimental parameters were tested. First with a square signal with frequency of the order of $10^4$ Hz: discharge duration is thus 50% of the period duration, that is 50μs. Second with a frequency of the order of $10^3$ Hz (frequency had to be reduced for technical reasons) and with a pulsed regimen of duration 300ns. In both cases the applied voltage and the inter-electrode distance were identical. The discharge duration seems to affect the vaporization of carbon. With longer pulse the quantity of amorphous carbon on the HOPG and the size of the amorphous carbon particles is higher. That's what lead to the pulsed regimen so as to limit the formation of amorphous carbon.

Intuitively, one can imagine that the energy brought to the electrodes is proportional to the pulse length and thus the longer the pulse, the more surface is heated over the vaporizing temperature of graphite. This phenomenon is counterproductive because longer pulse duration has also been linked to bigger particles. More precisely: with longer pulse the size distribution of the particles is very wide: lateral dimensions go from the micrometric scales up to millimetric scales; and thickness varies proportionally to lateral dimensions. For shorter pulse however, this size distribution is narrowed down toward the smaller sizes. It is supposed that there is an optimal pulse duration and associated frequency depending on the material and the experimental parameters to maximize exfoliation and minimize defects.

As far as the effectiveness of the process is concerned, high quality few layers' graphene was systematically observed during TEM characterization. Although some particles exhibited defects, there was also intact ones. As mentioned previously, no quantitative evaluation of the particles thickness was possible however, given the technical limitation, it can be affirmed that this process is an innovative and interesting way of producing few layers' graphene with lateral dimensions between 1 and $10^2$ μm.

Without any intermediate treatment of the exfoliated particles (ethanol dispersion, sonication or centrifugation), by only observing the deposited particles on the TEM grids, results were already encouraging. Had the best particles been isolated via such intermediate treatment, the TEM observation would not have been representative of all the products of such method.

CONCLUSIONS

This study further confirms the interest for nanoparticle production via plasma treatment. Although, the intrinsic mechanics and details behind the plasma formation and its role in exfoliation are not completely discovered. By tweaking different parameters such as the frequency, electrodes' orientation and pulse duration it was possible to point out the role of the different elements involved in such process. In general, three principal exfoliating forces have to be further investigated: thermal cycles, plasma erosion and shockwave formation. Also, plasma diagnosis and rapid imagery would greatly help decorrelating those aforementioned parameters.

ACKNOWLEDGEMENTS

Research was supported by the Institut Jean Lamour.REFERENCES

1. Novoselov, K. (2004). Electric Field Effect in Atomically Thin Carbon Films. *Science*, 306(5696), pp.666-669.
2. Pang, S., Englert, J., Tsao, H., Hernandez, Y., Hirsch, A., Feng, X. and Müllen, K. (2010). Extrinsic Corrugation-Assisted Mechanical Exfoliation of Monolayer Graphene. *Advanced Materials*, 22(47), pp.5374-5377.
3. Chang, Y., Kim, H., Lee, J. and Song, Y. (2010). Multilayered graphene efficiently formed by mechanical exfoliation for nonlinear saturable absorbers in fiber mode-locked lasers. *Applied Physics Letters*, 97(21), p.211102.
4. Balandin, A., Ghosh, S., Bao, W., Calizo, I., Teweldebrhan, D., Miao, F. and Lau, C. (2008). Superior Thermal Conductivity of Single-Layer Graphene. *Nano Letters*, 8(3), pp.902-907.
5. Lee, C., Wei, X., Kysar, J. and Hone, J. (2008). Measurement of the Elastic Properties and Intrinsic Strength of Monolayer Graphene. *Science*, 321(5887), pp.385-388.
6. Stoller, M., Park, S., Zhu, Y., An, J. and Ruoff, R. (2008). Graphene-Based Ultracapacitors. *Nano Letters*, 8(10), pp.3498-3502.
7. Nair, R., Blake, P., Grigorenko, A., Novoselov, K., Booth, T., Stauber, T., Peres, N. and Geim, A. (2008). Fine Structure Constant Defines Visual Transparency of Graphene. *Science*, 320(5881), pp.1308-1308.
8. Reina, A., Jia, X., Ho, J., Nezich, D., Son, H., Bulovic, V., Dresselhaus, M. and Kong, J. (2009). Large Area, Few-Layer Graphene Films on Arbitrary Substrates by Chemical Vapor Deposition. *Nano Letters*, 9(1), pp.30-35.
9. Guy, O., Burwell, G., Tehrani, Z., Castaing, A., Walker, K. and Doak, S. (2012). Graphene Nano-Biosensors for Detection of Cancer Risk. *Materials Science Forum*, 711, pp.246-252.
10. Van Noorden, R. (2011). Chemistry: The trials of new carbon. *Nature*, 469(7328), pp.14-16.
11. Chandra, V., Park, J., Chun, Y., Lee, J., Hwang, I. and Kim, K. (2010). Water-Dispersible Magnetite-Reduced Graphene Oxide Composites for Arsenic Removal. *ACS Nano*, 4(7), pp.3979-3986.
12. Lee, W., Park, J., Kim, Y., Kim, K., Hong, B. and Cho, K. (2011). Control of Graphene Field-Effect Transistors by Interfacial Hydrophobic Self-Assembled Monolayers. *Advanced Materials*, 23(30), pp.3460-3464.
13. Hazra A, D. (2014). Growth of Multilayer Graphene by Chemical Vapor Deposition (CVD) and Characterizations. *Journal of Nanomaterials & Molecular Nanotechnology*, s1.
14. Kim, K., Zhao, Y., Jang, H., Lee, S., Kim, J., Kim, K., Ahn, J., Kim, P., Choi, J. and Hong, B. (2009). Large-scale pattern growth of graphene films for stretchable transparent electrodes. *Nature*, 457(7230), pp.706-710.
15. Li, X., Cai, W., An, J., Kim, S., Nah, J., Yang, D., Piner, R., Velamakanni, A., Jung, I., Tutuc, E., Banerjee, S., Colombo, L. and Ruoff, R. (2009). Large-Area Synthesis of High-Quality and Uniform Graphene Films on Copper Foils. *Science*, 324(5932), pp.1312-1314.
16. Sun, Z., Yan, Z., Yao, J., Beitler, E., Zhu, Y. and Tour, J. (2010). Growth of graphene from solid carbon sources. *Nature*, 468(7323), pp.549-552.
17. Sutter, P., Flege, J. and Sutter, E. (2008). Epitaxial graphene on ruthenium. *Nature Materials*, 7(5), pp.406-411.
18. Cheng, M., Yang, R., Zhang, L., Shi, Z., Yang, W., Wang, D., Xie, G., Shi, D. and Zhang, G. (2012). Restoration of graphene from graphene oxide by defect repair. *Carbon*, 50(7), pp.2581-2587.
19. Nandamuri, G., Roumimov, S. and Solanki, R. (2010). Remote plasma assisted growth of graphene films. *Applied Physics Letters*, 96(15), p.154101.
20. Ostrikov, K., Neyts, E. and Meyyappan, M. (2013). Plasma nanoscience: from nano-solids in plasmas to nano-plasmas in solids. *Advances in Physics*, 62(2), pp.113-224.
21. Lee, H., Bratescu, M., Ueno, T. and Saito, N. (2014). Solution plasma exfoliation of graphene flakes from graphite electrodes. *RSC Adv.*, 4(93), pp.51758-51765.
22. Segundo, E., Fontana, L., Recco, A., Scholtz, J., Nespolo Vomstein, M. and Becker, D. (2018). Graphene nanosheets obtained through graphite powder exfoliation in pulsed underwater electrical discharge. *Materials Chemistry and Physics*, 217, pp.1-4.
23. Gao, X., Yokota, N., Oda, H., Tanaka, S., Hokamoto, K., Chen, P. and Xu, M. (2019). Preparation of Few-Layer Graphene by Pulsed Discharge in Graphite Micro-Flake Suspension. *Crystals*, 9(3), p.150.
24. Chae, S., Bratescu, M. and Saito, N. (2017). Synthesis of Few-Layer Graphene by Peeling Graphite Flakes via Electron Exchange in Solution Plasma. *The Journal of Physical Chemistry C*, 121(42), pp.23793-23802.
25. Subrahmanyam, K., Panchakarla, L., Govindaraj, A. and Rao, C. (2009). Simple Method of Preparing Graphene Flakes by an Arc-Discharge Method. *The Journal of Physical Chemistry C*, 113(11), pp.4257-4259.
26. Krivenko, A., Manzhos, R., Kotkin, A., Kochergin, V., Piven, N. and Manzhos, A. (2019). Production of few-layer graphene structures in different modes of electrochemical exfoliation of graphite by voltage13